\documentstyle[aps,epsf,psfig,multicol,rcs]{revtex}

\newcommand{\EF}{E_{\text{F}}}
\newcommand{\iint}{\int\!\!\!\int}
\newcommand{\ds}{\displaystyle}
\def\sgn{\mathop{\mbox{sgn}}\nolimits}
\newcommand{\nin}{\noindent}
\newcommand{\be}{\begin{equation}}
\newcommand{\ee}{\end{equation}}

\begin{document}
\draft

\title{Unbounded fluctuations in transport through an integrable cavity}
\author{Paul Pichaureau$^{1}$ and Rodolfo A. Jalabert$^{1,2}$}
\address{1:~Institut de Physique et Chimie des Mat\'eriaux de Strasbourg,
            23 rue du Loess, 67037 Strasbourg Cedex, France\\
         2:~Universit\'e Louis Pasteur, 3-5 rue de l'Universit\'e,
	    67084 Strasbourg Cedex, France}

\date{\today}

\maketitle
\widetext

\begin{abstract}
We derive a semiclassical scheme for the conductance through a rectangular cavity. 
The transmission amplitudes are expressed as a sum over families of trajectories 
rather than a sum over isolated trajectories. The contributing families are 
obtained from the evaluation of a finite number of continued fractions. We find
that, contrary to the chaotic case, the conductance fluctuations increase with 
the incoming energy and the correlation function exhibits a singularity at the
origin.
\end{abstract}

\pacs{PACS numbers:  72.23.Ad, 03.65.Sq}

\raggedcolumns
\begin{multicols}{2}
\narrowtext


\section {Introduction}
\label{sec:Intro}

The density of states of small quantum systems is known to be sensitive to
the underlying classical mechanics. The chaotic or integrable character of
the classical dynamics translates into the statistical properties of the
spectra \cite{RevBoh}. The spectral correlations of classically chaotic
systems are described according to random matrix theory \cite{BGS} and the
fluctuations with respect to the mean density of states are less pronounced than
for integrable systems. The connection between classical and quantum properties
is particularly clear within the semiclassical approximation: the density of
states of classically chaotic systems is expressed in the Gutzwiller trace
formula \cite{gutz90} as a sum over [isolated] periodic orbits, while
integrable systems are described by the Berry-Tabor formula \cite{ber77}, where
the the density of states is given as a sum over families of trajectories
(or invariant tori). 

What is the corresponding situation in open systems? Transport properties
instead of spectral ones are usually considered. The transmission amplitudes
and probabilities are given as semiclassical expansions over trajectories
that traverse the scattering region \cite{Miller,jal}, providing a link 
between classical and quantum properties. Such an approach is firmly established
in the case of isolated trajectories, as typically found for chaotic dynamics,
but only recently the subtleties associated with an integrable geometry have
started to be addressed \cite{Chaos,ishio95,Lin,schwi96,Wirtz}. As in the
Berry-Tabor formula for the density of states, the existence of families of
trajectories degenerate in action modifies the structure of the semiclassical
expansion. Integrable cavities fed by leads present the additional 
problem that conserved quantities in the scattering region are not necessarily
conserved in the leads. The semiclassical description of transport through
integrable cavities is a necessary step towards understanding the influence
of the underlying classical mechanics in open systems. This is the goal of this work, where
we present a semiclassical evaluation of the transmission amplitude through a
rectangular cavity and apply it to the study of the conductance fluctuations.

The interest in open systems stems from their parallel with closed systems
and also from the physically measurable realizations that have been developed in 
the last few years. High mobility-semiconductor microstructures at low
temperatures give access to the ballistic regime, where transport is dominated
by the geometrical scattering of electrons off the walls of the lithographically
defined cavities \cite{marcu,timp,csf}. On the other hand, microwave cavities provide 
an easily tunable system to study the propagation of electromagnetic waves in a 
defect-free region \cite{stoec}.

Quantum interference in ballistic cavities gives rise to conductance
fluctuations under a small perturbation of the system (magnetic field,
Fermi energy of the incoming electrons, shape of the cavity, etc)
\cite{marcu,keller,lee97}. For chaotic cavities, the characteristic 
correlation lengths of the fluctuations are well described by the existing
semiclassical theories \cite{blusm,jal}, while the universal character of the 
variance has been addressed within a Dyson hypothesis for the scattering matrix
\cite{meba,ropibe2}. Random matrix theories are obviously not 
appropriate to treat integrable systems, forcing us to rely on semiclassical
expansions and numerical calculations. The early experiments on ballistic
cavities \cite{marcu} yielded a more structured spectrum of conductance fluctuations
for the integrable case than for the chaotic one. This tendency was reproduced
by numerical calculations \cite{Chaos} showing that in rectangular cavities
the higher harmonics of the conductance spectrum are more pronounced than in 
the chaotic case. Also, the variance of the conductance appeared as increasing
with the mode number. The lack of transport theories for integrable systems 
(which do not enjoy the universality properties of chaotic cavities) and the 
difficulties with their fabrication are probably responsible
for the fact that subsequent efforts on conductance fluctuations were mainly 
centered on the chaotic case \cite{huibers}.

Weak-localization, the decrease of the average conductance due to constructive
interference of time-reversed backscattering trajectories, is another quantum
interference effect which has recently been studied in ballistic cavities. The
underlying classical mechanics was proposed \cite{bar} to translate into a 
different shape of the magnetoconductance peaks: a lorentzian line-shape for
chaotic cavities and a triangular form for the integrable case. Chaotic 
cavities indeed exhibited the lorentzian line-shape while rectangular and circular 
geometries showed triangular or lorentzian peaks depending on the experiment 
\cite{keller,lee97,chang94,berry94,lutj}. The importance of the leads has been demonstrated
\cite{bird} and the disagreements between different experiments are not settled yet. 
This illustrates the difficulties in dealing with integrable cavities and the importance 
of developing appropriate theoretical descriptions.

In this work we consider a square cavity (of length side $1$) with hard walls and
connected to leads on opposite sides of the square by openings of size
$W$ (see inset Fig.~\ref{fig:I}). (For our purposes square cavities are no less 
general than rectangular ones since we can always scale one of the sides.)
We develop in Sec.~\ref{sec:Semi} 
a semiclassical expansion of the transmission amplitude between two modes
in terms of families of trajectories degenerate in action. The determination
of these families (Sec.~\ref{sec:Cofra}) results in a diophantic problem and
the evaluation of a finite number of continued fractions. This constitutes
a very efficient numerical method that is to be compared with the numerical
quantum approaches based on discretization of real space or truncation of
Hilbert space. The classical conductance (Sec.~\ref{sec:meco}) and
the unbounded fluctuations (Sec.~\ref{sec:coflu}), increasing with the 
incoming energy, are obtained from the numerical implementation of the
continued fraction approach. The correlation function of the conductance,
when properly normalized, is shown to be well defined in all the energy range
and shows a cusp-like singularity at the origin. Further approximations on 
our semiclassical expansions allow us to demonstrate that the unbounded
conductance fluctuations grow linearly with the incoming Fermi wave-vector
$k$. In the conclusions (Sec.~\ref{sec:conclu}) we analyze the experimental
relevance of our findings and the implications for other integrable geometries.

Transport through a square cavity has recently been considered by Wirtz and
collaborators \cite{Wirtz}, where alternative forms of the transmission
coefficient were proposed and the correspondence between quantum results and
classical trajectories was established. The dynamics through rectangular
cavities constitutes a paradigm for the interplay between scattering and
integrability, and has been treated in other contexts. For instance, it
has been presented by Zwanzig \cite{zwanzig} as an example where the
short memory approximation to the migration of a classical dynamical system
between regions of configuration space is entirely wrong. 


\section {Semiclassical formulation}
\label{sec:Semi}

We consider a phase-coherent cavity connected to two reservoirs through
leads that support $N$ propagating modes.
The conductance in the Landauer formalism is 
proportional to the transmission coefficient $T$ 

\be
  \label{eq:land}
   g = \frac{e^2}{h} \ T=\frac{e^2}{h} \sum_{a,b=1}^{N} |t_{ba}|^2 \ .
\ee

\nin The complex numbers $t_{ba}$ are the transmission amplitudes from the 
mode $a$ at the entrance lead to the mode $b$ at the exit lead.  They 
are related to the Green function (evaluated at the Fermi energy 
$\EF$) connecting the point $(0,y)$ at the entrance with $(L,y')$ at 
the exiting lead through \cite{fish81}:
\begin{eqnarray}
  \label{eq:tabG}
  t_{ba} = & \ds -i\hbar \sqrt{v_a v_b} \iint dy\;dy'\; \phi_b^*(y') \,
  \phi_a(y) \nonumber \\ 
  & \lefteqn{ {} \times G(L,y';0,y\,;\EF)} \ ,
\end{eqnarray}
where we have discarded an unimportant phase factor. $\phi_{a,b}$ 
represent the transverse wave functions in the leads of width $W$
($\phi_a(y) = \sqrt{2/W} \sin(a\pi y/W)$) and $v_{a,b}$ the longitudinal
velocities associated with the modes $a$ and $b$.

In the semiclassical approximation, the Green function is expressed
as a sum over trajectories joining the points $(0,y)$ and $(L,y')$ 
\cite{gutz90}:
\begin{eqnarray}
  \label{eq:Gscl}
  G_{\text{scl}}(L,y';0,y;\EF) = & \ds \frac{2\pi}{(2\pi i
    \hbar)^{3/2}}
  \sum_{s(y,y')} \sqrt{D_s} \nonumber\\
  &  {}\times \exp \left[\frac{i}{\hbar} S_s (y'\!,y,\EF) -
    i\frac{\pi}{2}\mu_s \right] \ .
\end{eqnarray}
$S_s$ is the classical action along the trajectory $s$. Since we
are considering billiards without magnetic field $S_s=k L_s$, with
$k=mv/\hbar$ the Fermi wavevector and $L_s$ the trajectory length.
Denoting $\theta$ and $\theta'$ the incoming and outgoing angles,
the preexponential factor is given by $D = (v\cos{\theta'}/m)^{-1}
|(\partial \theta / \partial y')_y|$. We include in the phase $\mu_s$ 
the Maslov index counting the number of constant-energy conjugate points and
the phase acquired at the bounces with the hard walls.

The semiclassical expression of the transmission amplitude is 
obtained by using the above approximation of the Green function in 
Eq.~(\ref{eq:tabG}) and performing the $y$ and $y'$ integrals to 
leading order in $\hbar$. A stationary-phase integration over $y$ 
selects the trajectories entering the cavity with the angle 
$\theta_{\bar{a}}$ ($\bar{a}=\pm a$ and $\sin\theta_{\bar{a}} = 
\bar{a}\pi/ kW$) and leaving at the point $y'$, leading to
\begin{eqnarray}
  \label{eq:tbaSPA1}
t_{ba} = & \ds i \ \sqrt{\frac{v_b}{2W}} \int d y' \; \phi_b(y') \sum_{\bar{a}=\pm a} 
\sum_{s(\theta_{\bar{a}},y')} \sgn \bar{a} \sqrt{B_{s}} \nonumber \\
& {}\times \exp \left[ \frac{i}{\hbar} \tilde{S}_{s} (y',\theta_{\bar{a}},\EF) - 
i\frac{\pi}{2}\tilde{\mu}_s \right].
\end{eqnarray}
The reduced action is
\begin{equation}
  \label{eq:redact}
  \tilde{S}(y',\theta_{\bar{a}},\EF) = S(y'\!,y_0,\EF) +\hbar\pi\bar{a}y_0/W \ ,
\end{equation}
where $y_0(\theta_{\bar{a}},y')$ is the initial point selected from the
stationary-phase integration. The prefactor is now given by
$B=(v\cos{\theta'})^{-1} |(\partial y /\partial y')_{\theta}|$,
and the Maslov index is increased by one if $(\partial \theta /
\partial y)_{y'}$ is positive. 

\begin{figure}[tbh]
\setlength{\unitlength}{1mm}
\begin{picture}(70,93)(10,10)
\put(0,0){\epsfxsize=10cm\epsfbox{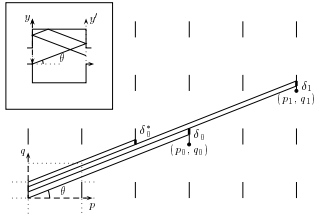}}
\end{picture}
\vspace{-4cm}
\caption{Unfolded space for the dynamics in a square cavity (inset).
The trajectory entering at the lowest point ($y\!=\!0$) of the left
lead with an angle $\theta$ is shown in the original and extended
space. It belongs to the family that leaves the cavity through the
exiting lead $(p_0,q_0)$ and has a weight $\delta_0$. For the
particular chosen $\theta$ there are two families contributing
to the transmission amplitude (0 and 1) and one family contributing
to the reflection amplitude ($0^{*}$).}
\label{fig:I}
\end{figure}

When nearby trajectories are non-degenerate in action, as is typically
the case in chaotic cavities, the integral over $y'$ can also be done
by stationary-phase and the semiclassical $t_{ba}$ is expressed as a
sum over trajectories with quantized initial and final angles
\cite{jal,Chaos}. On the other hand, when trajectories come in families
with the same action (or length) the phase in Eq.~(\ref{eq:tbaSPA1}) is
linear in $y'$, preventing another stationary-phase integration. This is
the case of the trajectories directly crossing a ballistic cavity from
the entrance to the exiting lead, where the $y'$-integral can be
exactly done \cite{Chaos,Lin}. The trajectories going through a square
cavity also come in families, or bundles \cite{Wirtz}, and the 
integration over the $y'$-coordinate is very similar to the case of 
direct trajectories since the dynamics in an extended space is that of
free particles (Fig.~\ref{fig:I}).

Symmetry arguments for a rectangular cavity dictate that $t_{ba}\!=\!0$ if
$a\!+\!b$ is odd. For even $a\!+\!b$ we perform the $y'$-integration for
each family $n$ defined by the coordinates $(p_n,q_n)$ of the
exiting lead in the extended space obtaining
\begin{eqnarray}
  \label{eq:tabinterr}
  t_{ba}= -\frac{i}{W} \ \sqrt\frac{\cos\theta_b}{\cos\theta_a}
  \sum_{n} \left\{I_{n}(a\!+\!b)-I_{n}(a\!-\!b)\right\} \nonumber \\
  \times \exp{\left[i \left(k L_n\!+\!\frac{\pi a}{W} \left(q_n\!-\!L_n 
  \sin{\theta_a}\!+\!W \varepsilon(q_n)\right)\!-\!\frac{\pi}{2} 
  \tilde{\mu}_n\right)\right]}
\end{eqnarray}
where 
\begin{eqnarray}
  \label{eq:idef}
I_{n}(x) = \frac{W}{\pi x} \left(\exp{\left[\frac{i\pi x}{W} 
y_{f}^{\prime(n)}\right]} - \exp{\left[\frac{i\pi x}{W} y_{i}^{\prime(n)}\right]}
\right) \ .
\end{eqnarray}

\nin $y_{i}^{\prime(n)}$ and $y_{f}^{\prime(n)}$ are the extreme points of
the exiting interval, $\varepsilon(q_n)\!=\!0$ for even $q_n$ and 
$\varepsilon(q_n)\!=\!1$ for odd $q_n$. The parity of $q_n$ appears in 
the phase due to mirror symmetries involved in going to the extended
space. The trajectories of the $n$-th family have a length 
$L_n = p_n/\cos{\theta_a}$ (all lengths are expressed in units of the 
side of the square). In the extended space we have free motion, therefore 
the phase $\tilde{\mu}_n$ is simply given by the $p_n\!-\!1$ bounces with the 
vertical walls and the $q_n$ bounces with the horizontal ones; that is, 
$\tilde{\mu}_n=2(p_n\!+\!q_n\!-\!1)$.
The trajectories that contribute to the transmission amplitude are those 
going from the left to the right lead, therefore only the values $n$ with
odd $p_n$ should be considered. Conversely, the even values of $p_n$ 
yield the reflection amplitude. For transmission amplitude we can simplify
the phase of Eq.~(\ref{eq:tabinterr}) and write

\begin{eqnarray}
  \label{eq:tabinter}
  t_{ba}= -\frac{i}{W} \ \sqrt\frac{\cos\theta_b}{\cos\theta_a}
  \sum_{n}  \varepsilon_n \exp[i k \tilde{L}_n] \nonumber \\
  {}\times \left\{I_{n}(a\!+\!b)-I_{n}(a\!-\!b)\right\} \ .
\end{eqnarray}

\nin We have defined the phase

\be
\label{eq:epsilonsn}
\varepsilon_n=\exp[i\pi(a\!+\!1) \varepsilon(q_n)] \ ,
\ee

\nin and the reduced length

\be
\label{eq:redlength}
\tilde{L}_n = p_n \cos{\theta_a} + q_n \sin{\theta_{a}} \ .
\ee

Similarly to the case of direct trajectories \cite{Chaos},
$\bar{a}\!+\!\bar{b}=0$ has to be treated separately for the $y'$-integration.
However, the corresponding result is included in Eq.~(\ref{eq:tabinter}) by taking the
limit $x \! \rightarrow \! 0$. Obviously, $\bar{a}\!+\!\bar{b}=0$ corresponds to the
maximum transmission since this is the case where the classical
trajectory arrives to the exiting lead with the quantized angle of mode
$b$. As emphasized in Ref.~\cite{Wirtz}, the diffractive terms obtained
for $\bar{a}\!+\!\bar{b} \neq 0$ become more important for families with
a weight $\delta_n=y_{f}^{\prime(n)}\!-\!y_{i}^{\prime(n)} \ll W$.


\section {Continued fraction representation}
\label{sec:Cofra}

In order to calculate the semiclassical transmission amplitude of 
Eq.~(\ref{eq:tabinter}) we need to determine all families $n$, with
their exiting lead $(p_n,q_n)$ and extreme points $y_{f,i}^{\prime(n)}$.
This evaluation naturally leads to a diophantic problem \cite{khinchin}.
As we show below, the coefficients $(p_n,q_n)$ are part of the 
intermediate fractions (or Farey series) appearing in the continued fraction
representation of $1/\tan{\theta}_{\bar{a}}$. 

For a given angle $\theta$ and an initial point $y_0$, the trajectory
in extended space is the straight line $D(y_0)$ defined by
$y=y_0\!+\!x\tan{\theta}$. Let us start with the trajectory entering the 
cavity at the lowest point ($y_0\!=\!0$), whose exiting lead is defined by the 
segment $[(p_0,q_0);(p_0,q_0+\!W)]$ which intersects $D(0)$ (see Fig.~\ref{fig:I}). 
As we increase $y_0$, we will remain within a family of degenerate trajectories 
(that we note by $n\!=\!0$) until the exiting point $y_0\!+\!p_0 \tan{\theta}$ hits the 
uppermost point of the segment. The pair $(p_0,q_0)$ must verify the conditions:

\begin{enumerate}
\item  [{\em a.}] \ $0<p_0\tan\theta -q_0<W$ \ ,

\item  [{\em b.}] \ $\forall (p,q) \ {\rm such} \ {\rm that} \ 0 < p\tan\theta -q < p_0\tan\theta -q_0 
\Rightarrow p > p_0 $ .
\end{enumerate}
According to $a$, the first $y$-interval is $[0,\delta_0]$ (or equivalently,
the first $y'$-interval is $[W-\delta_0,W]$), while condition $b$ means that
$(p_0,q_0)$ is the first lattice point verifying $a$. The uppermost family will be
associated with an interval $[(p^*_0,q^*_0);(p^*_0,q^*_0\!+\!W)]$, where the pair
$(p^*_0,q^*_0)$ verifies similar conditions as $(p_0,q_0)$:
\begin{enumerate} \setcounter{enumi}{2}
\item  [{\em c.}] \ $0<W+p^*_0\tan\theta-q^*_0<W$ \ ,
\item  [{\em d.}] \ $\forall (p,q) \ {\rm such} \ {\rm that} \ W\!+\!p^*_0\tan\theta\!-\!q^*_0<
W\!+\!p\tan\theta\!-\!q<W \Rightarrow 
p>p^*_0$ \ .
\end{enumerate}
According to $c$, the first $y$-interval of the uppermost family is
$[W\!-\!\delta_0^*,W]$ (the first $y'$-interval is $[0,\delta_0^*]$),
while $d$ implies that $(p^*_0,q^*_0)$ is the first lattice point 
verifying $c$.

Now that we have determined the lowest and the uppermost families for the
$[0,W]$ $y$-interval, the following sequences of families can be obtained
by reducing ourselves to the $y$-interval $[\delta_0,W\!-\!\delta_0^*]$, and with
the changes of $(p_0,q_0)$, $(p^*_0,q^*_0)$ by $(p_1,q_1)$, $(p^*_1,q^*_1)$
the conditions $a-d$ define the next two families. Continuing this procedure
until the two sequences of families begin to overlap each other, we obtain
all the terms to be included in the sum of Eq.~(\ref{eq:tabinter}). For the 
sequence of lower families we have

   \begin{mathletters}
        \label{allexl}
        \begin{eqnarray}
y'_{i}(p_l,q_l)= W+(p_l-p_{l-1})\tan\theta-(q_l-q_{l-1}) \ ,
	\label{exla} 
	\end{eqnarray}
        \begin{eqnarray}
y'_{f}(p_l,q_l)= W \ ,
        \label{exlb}
        \end{eqnarray}
        \begin{eqnarray}
\delta_l=y'_{f}-y'_{i}=q_l-q_{l-1}-(p_l-p_{l-1})\tan\theta \ ,
        \label{exlc}
        \end{eqnarray}
   \end{mathletters}

\nin while for the upper-families we have

   \begin{mathletters}
        \label{allexu}
        \begin{eqnarray}
y'_{i}(p_u^*,q_u^*)= 0 \ ,
        \label{exua}
        \end{eqnarray}
        \begin{eqnarray}
y'_{f}(p_l^*,q_l^*)= (p_u^*-p_{u-1}^*)\tan\theta-(q_u^*-q_{u-1}^*) \ ,  
        \label{exub}
        \end{eqnarray}
        \begin{eqnarray}
\delta_u^*=(p_u^*-p_{u-1}^*)\tan\theta-(q_u^*-q_{u-1}^*) \ .
        \label{exuc}
        \end{eqnarray}
   \end{mathletters}

\nin The very last family is simultaneously shadowed by lower and upper 
families, therefore has $y'_{i}$ given by (\ref{exla}) and $y'_{f}$ by
(\ref{exub}).

We can now establish the relationship with the continued fraction representation of 
$\Theta\!=\!1/\tan\theta$, that is defined by the sequences $(\alpha_m)$ and 
$(a_m)$ as follows \cite{khinchin}:
\begin{equation}
\label{eq:khin1}
\begin{array}{cc}
  \alpha_0 = \Theta & a_0 = [\alpha_0] \\
  \alpha_{m+1} = \ds \frac1{\alpha_m - a_m} &  \quad a_{m+1} =
  [\alpha_{m+1}] \ .
\end{array}
\end{equation}

\nin $[\alpha]$ denotes the integer part of $\alpha$. The best rational
approximations to $\Theta$ are the fractions $P_{m}/Q_{m}$, called
convergents, and obtained from the recurrence relations
\begin{equation}
\label{eq:khin2}
\left\{\begin{array}{ll}
      P_{m}=P_{m-2}+a_m P_{m-1} \\
      Q_{m}=Q_{m-2}+a_m Q_{m-1} \ , 
    \end{array}
\right.
\end{equation}

\nin with the initial choice of $(P_{-1},Q_{-1})=(1,0)$ and
$(P_{0},Q_{0})=([\Theta],1)$. Since $(P_m)$ and $(Q_m)$ are sequences of
integers, we can represent the convergents as lattice points
$(P_m,Q_m)$ that approach the straight line $D(0)$ as $m$ increases from
above (even $m$) and below (odd $m$). Moreover, the convergents verify
\begin{enumerate}
\item  [{\em e.}] \ $\forall (p,q) \ {\rm such} \ {\rm that} \ |p\tan\theta -q| <
|P_m\tan \theta - Q_m|
\Rightarrow p > P_m $.
\end{enumerate}

The intermediate lattice points on the segment
$[(P_{m-2},Q_{m-2}),(P_{m},Q_{m})])$ define the $m$-th Farey sequence
(or intermediate fractions) by
\begin{equation}
\label{eq:deffar}
\left\{
  \begin{array}[c]{l}
    p_{m}^k = P_{m-2} + k  P_{m-1}\\
    q_{m}^k = Q_{m-2} + k  Q_{m-1}
  \end{array}
\right. \quad 0\leq k \leq a_m \ .
\end{equation}

\nin The sequence of equally spaced lattice points $(p_{m}^k,q_{m}^k)$
starts at the convergent $(P_{m-2},Q_{m-2})$ (for $k\!=\!0$) and finishes at
$(P_{m},Q_{m})$ (for $k\!=\!a_m$). The translation vector is given by the
coordinates of the convergent $(P_{m-1},Q_{m-1})$ and the intermediate
fractions verify the properties
\begin{enumerate}
\item  [{\em f.}] \ $\forall m,k,(p,q) \ {\rm such} \ {\rm that} \ 0<p\tan\theta -q<p_{2m+1}^k\tan \theta -
q_{2m+1}^k \Rightarrow p>p_{2m+1}^k$ \ ,

\item  [{\em g.}] \ $\forall m,k,(p,q) \ {\rm such} \ {\rm that} \ q_{2m}^k-p_{2m}^k\tan\theta>q-p\tan\theta>0
\Rightarrow p>p_{2m}^k$ \ . 

\end{enumerate}

These are the conditions $b$ and $d$, respectively, determining the lattice points 
associated with the exiting leads. Therefore, the two types
of families contributing to $t_{ba}$ are given by the
Farey sequences of even and odd order convergents, with the the
additional requirements that $p_n$ is odd and that the exiting points are closer than 
$W$ to the straight line $D(0)$.

The continued fraction representation of a generic (irrational) real number 
results in an infinite sequence of convergents. However, the fact that 
we have defined a finite distance $W$ to approach $D(0)$ implies that once the
lower and upper families begin to overlap the sum in Eq.~(\ref{eq:tabinter}) can
be cut. Therefore, for each angle $\theta_a$ the sum is actually finite.
In addition, we show in the sequel that at most three sequences of intermediate fractions
contribute to it. Let us consider the first convergent $(P_m,Q_m)$ contributing
to the sum (the previous convergents being farther away than $W$ to $D(0)$), and assume 
for definiteness $m$ to be odd. Clearly, before this
convergent we can only have one contributing Farey sequence:

\be
\label{eq:ffs}
(p_{m}^k,q_{m}^k) \hspace{1cm} {\rm with} \hspace{0.5cm} k \in (1,a_{m}) \ .
\ee

\nin The $k\!=\!0$ case is excluded because otherwise $(P_m,Q_m)$ would not be
the first convergent of the sum. The following sequence,

\be
\label{eq:sfs}
(p_{m+1}^k,q_{m+1}^k) \hspace{1cm} {\rm with} \hspace{0.5cm} k \in (1,a_{m+1}) \ ,
\ee

\nin is relevant since it finishes at the next convergent $(P_{m+1},Q_{m+1})=
(p_{m+1}^{m+1},q_{m+1}^{m+1})$, which must contribute to the sum. It is closer to
$D(0)$ than $(P_m,Q_m)$) and it is not completely shadowed by $(P_m,Q_m)$ 
(they are at opposite sides of $D(0)$). Consequently, there are two
possibilities: i) $(P_{m+1},Q_{m+1})$ is partially shadowed by $(P_m,Q_m)$
and therefore the sum stops there since the points $(p_{m+2}^k,q_{m+2}^k)$
with $k\!>\!0$ will be completely shadowed; ii) $(P_{m+1},Q_{m+1})$ is not
shadowed by $(P_m,Q_m)$, therefore $(p_{m+2}^1,q_{m+2}^1)$ is 
simultaneously shadowed by $(P_m,Q_m)$ and $(P_{m+1},Q_{m+1})$ since
$W\!+\!q_{m+2}^1\!-\!p_{m+2}^1 \tan{\theta} > Q_{m+1}\!-\!P_{m+1} \tan{\theta} > 0$
is equivalent to $W\!+\!Q_{m}\!-\!P_{m}\tan{\theta} > 0$, which is precisely
the condition for $(P_m,Q_m)$ being an exiting point. The points 
$(p_{m+2}^k,q_{m+2}^k)$ with $k\!>\!1$ are completely shadowed, and therefore
the series stops there. In conclusion, in the case i) the contributing
Farey sequences are those of Eqs.~(\ref{eq:ffs})-(\ref{eq:sfs}), while in case ii) we have
to add one intermediate fraction of the next family, namely $(p_{m+2}^1,q_{m+2}^1)$.

The above considerations simplify the semiclassical transmission
amplitude of Eq.~(\ref{eq:tabinter}) to a finite sum, having at most three
sequences of intermediate fractions. In case i) the form of the $y'$-intervals
(Eqs.~(\ref{allexl})-(\ref{allexu})) is $[W\!-\!\delta_l,W]$ or $[0,\delta_u^*]$, 
therefore Eq.~(\ref{eq:tabinter}) can be written as

\begin{eqnarray}
  \label{eq:tab}
  t_{ba}=& \ds \frac{1}{W} \ \sqrt\frac{\cos\theta_b}{\cos\theta_a}
  \sum_{n=l,u} \varepsilon_n \varepsilon_n^{\prime} \exp{[ik \tilde{L}_n]} \nonumber \\
  & {}\times \left\{\Delta_n(a+b)-\Delta_n(a-b)\right\} \delta_n \ ,
\end{eqnarray}
\nin with the family-dependent function $\Delta_n$ defined by
\begin{equation}
  \label{eq:Delta}
   \Delta_n(x) = \frac{2W}{\pi x \delta_n} 
   \exp \left[{i\frac{\pi x \delta_n}{2W}}\right] \ \sin \left[\frac{\pi x
   \delta_n}{2W}\right] \ ,
\end{equation}

\nin $\varepsilon_n^{\prime}\!=\!1$ if $n\!=u$ (upper family) and 
$\varepsilon_n^{\prime}\!=\!-\!1$ if $n\!=l$ (lower family). 
As previously stated, only odd $p_n$ should be considered in the
expansion for the transmission amplitude. In case ii) the last family
does not have a $y'$-interval with the simple form of the previous ones.
Therefore, Eq.~(\ref{eq:tab}) depending only on the
weights $\delta_n$ can be used for all terms (families of trajectories),
except for the last one, were the form (\ref{eq:tabinter}) together with
the limits (\ref{exla}) and (\ref{exub}) should be employed. However, for
the last family the weight $\delta(p_m^1,q_m^1)$ is very small and to first
order in $(a \pm b)\delta(p_m^1,q_m^1)/W$, we can still use Eq.~(\ref{eq:tab}).
We stress that the subindex $n$ runs over contributing families, while $m$
labels the convergents (contributing or not) and the superindex $k$ orders the
intermediate fractions of the Farey sequences.

The forms (\ref{eq:tabinter}) and (\ref{eq:tab}) of the transmission amplitude 
provide a very
powerful method to numerically compute the conductance through a rectangular
cavity. For each $a\!=\!1\ldots N$, we only need to calculate a finite number of 
convergents and intermediate fractions of the continued fraction representation 
of $1/\tan{\theta_a}$ and recursively obtain the weighting intervals 
$\delta_n$. The advantage over quantum methods based on discretization
(recursive Green function, wave function matching, etc) is that it can be
used for large wavevectors $k$. The method actually gets more exact with
increasing $k$ since it involves a semiclassical approximation. Obviously,
Eq.~(\ref{eq:tab}) does not incorporate diffractive effects like reflection
off the lead mouths \cite{ishio95,schwi96,ingold}. However, as shown in 
Ref.~\cite{schwi96}, the semiclassical approach can be adapted in order to describe 
these simple diffractive effects.

\begin{figure}[tbh]
\setlength{\unitlength}{1mm}
\begin{picture}(10,53)(85,10)
\put(0,0){\epsfxsize=7cm\epsfbox{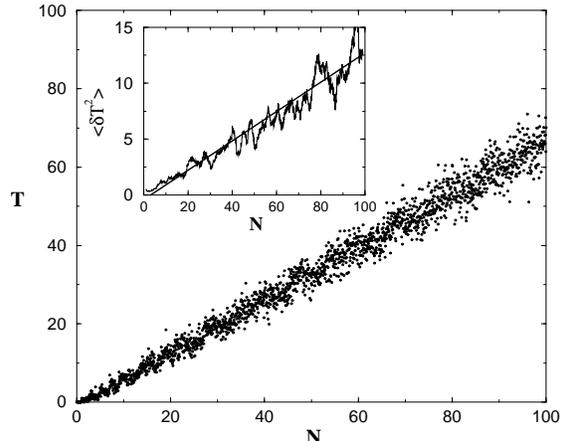}}
\end{picture}
\vspace{4mm}
\caption{Conductance through a square cavity in units of $e^2/h$
as a function of the number of modes $N=kW/\pi$ for an opening of
$W=0.2$ in units of the size of the square. Inset: local average
of the conductance fluctuations $\langle \delta T^2 \rangle$ versus
$N$. The straight line is the local variance of the fluctuations
computed from Eq.~(\protect\ref{eq:flufluf}).}
\label{fig:II}
\end{figure}

In Fig.~\ref{fig:II} we present the conductance of a square cavity calculated
from Eq.~(\ref{eq:tab}) for an opening $W\!=\!0.2$ in a large 
$k$-interval (the $k$-mesh shown is
relatively sparse for viewing purposes). We remark two salient features:
the linear increase of the mean conductance with $N=kW/\pi$ and the fluctuations
around the mean, which become larger as $k$ increases. The linear
increase of $\langle g \rangle$ with a slope given by the classical 
transmission coefficient constitutes a checking of the procedure, the 
increase of the fluctuations obtained within our semiclassical approach is
consistent with previous quantum computations \cite{Chaos}. In the following section
we show that further approximations on the semiclassical transmission amplitudes
$t_{ba}$ allow us to obtain both these features analytically.

As it has been shown in previous studies comparing semiclassical and quantum
results \cite{ishio95,schwi96,Wirtz}, the former reproduce the gross structure
of the fluctuations in the intermediate $k$-regime (not too small $k$ in order
for semiclassics to apply and not too large $k$ to prevent the quantum calculations
becoming unreliable). The quantum-semiclassical correspondence is particularly
clear at the level of the Fourier components of the transmission coefficients.
Specifically, for square cavities Wirtz {\em et al.} \cite{Wirtz} have
identified the peaks of the Fourier transforms with the families (or bundles)
of classical trajectories contributing in a semiclassical expansion. Unitarity,
the mathematical translation of charge conservation, is a critical test for 
semiclassical approximations. In our case, the difference $T+R-N$ grows with 
$N$, but it remains smaller than the fluctuations of $T$.


\section {Mean Conductance}
\label{sec:meco}

The semiclassical expression (\ref{eq:tab}) for the transmission amplitude is
very useful since, as we have shown in the previous chapters, the numerical
evaluation of a few continued fractions allows the calculation of the conductance.
In the sequel we further simplify this semiclassical approach in order to render
it analytically tractable.
The function $\Delta_n(x)$ defined in Eq.~(\ref{eq:Delta}) is peaked for $x\!=\!0$ 
(when the quantized angles of the incoming and outgoing modes coincide with the
angle of the trajectory) and decays on the scale of $W/\delta_n$, therefore it 
can be approximated by the rectangular function 
$$
\Pi_n(x) = \left\{  
 \begin{array}[c]{l}
    \ds \exp{\left[i\frac{\pi x \delta_n}{2W}\right]} \ , \qquad 
    \text{if} \quad |x|<\frac{W}{2\delta_n} \ , \\
    0 \qquad \text{otherwise \ .}
  \end{array}
\right.
$$
Thus, the semiclassical $t_{ba}$ simplifies to:

\be
\label{eq:tab2}
t_{ba} = \frac{1}{W}\sqrt\frac{\cos\theta_b}{\cos\theta_a}
\sum_{n} \Pi_n(a-b) \varepsilon_n \varepsilon_n^{\prime} \delta_n
\ \exp[ik \tilde{L}_n] \ .
\ee

The total transmission coefficient is obtained by summing the magnitude 
squared of the transmission amplitudes over the propagating modes. Therefore,
within our semiclassical approximation, it will be expressed as a sum 
over pairs of families of trajectories (with odd $p_n$ and $p_{n'}$).

\begin{eqnarray}
  \label{eq:simp}
  T=& \ds \frac{1}{W^2} \sum_{a,b=1}^{N}
  \frac{\cos\theta_b}{\cos\theta_a} \sum_{n,n'} \varepsilon_n 
  \overline{\varepsilon}_{n'} 
  \varepsilon_n^{\prime} {\varepsilon}_{n'}^{\prime}  \delta_n
  \delta_{n'} \nonumber \\
  & \ds {} \times \exp{[ik(\tilde{L}_n\!-\!\tilde{L}_{n'})]} \ 
  \Pi_n(a\!-\!b) \overline{\Pi}_{n'}(a\!-\!b) \ .
\end{eqnarray}

From this highly oscillating function of $k$, we want to extract its 
secular behavior, linear in $k$. Averaging $T(k)/k$ over all $k$ \cite{Chaos}
(or $T(k)$ over several oscillations) amounts to make the diagonal
approximation between the families of trajectories ($n\!=\!n'$). In the
well-studied case of isolated trajectories, the diagonal approximation
yields the classical probability of transmission by
pairing individual trajectories. In the present case, the concept of 
families of trajectories replaces the role of individual trajectories. 
Inserting the definition of the function $\Pi$, we find 
\begin{equation}
  \langle T\rangle= \frac{1}{W^2} \ \sum_{a=1}^{N} \ \sum_{n} \delta_n^2 
  \left( \sum_{b=b_{min}}^{b_{max}}
  \frac{\cos\theta_b}{\cos\theta_a} \right) \ ,
\end{equation}

\nin $b_{min}~=~\max{\{a\!-\!W/2\delta_n,1\}}$ and 
$b_{max}=\min{\{a\!+\!W/2\delta_n,N\}}$.
We now split the sum over families according to 

\begin{eqnarray}
\label{eq:simpspl}
\langle T\rangle= \frac{1}{W^2} \ \sum_{a=1}^{N} \frac{1}{\cos{\theta_a}}
\left[\sum_{0<\delta_n<\delta_{\alpha}} \delta_n^2 \sum_{b=1}^{N} 
\cos{\theta_b} + \right.
\nonumber \\
\left. \sum_{\delta_{\alpha}<\delta_n<\delta_{\beta}} \delta_n^2 
\sum_{b=b_{min}}^{b_{max}} \cos{\theta_b} +
\sum_{\delta_n>\delta_{\beta}} \delta_n^2 \sum_{b=a-W/2\delta_n}^{a+W/2\delta_n} 
\cos{\theta_b} \right] 
\end{eqnarray}

\nin with $\delta_{\alpha}$ and $\delta_{\beta}$ respectively the min and 
max of $\{W/2a,W/2(N-a)\}$. In the classical limit of $N=kW/\pi\gg 1$ the
last term dominates and the sum over $b$ can be approximated by an integral
leading to 

\begin{eqnarray}
\label{eq:simpspl2}
  \langle T \rangle  & = \ds \sum_{a=1}^{N} \sum_{n} \frac{\delta_n}{W} \ . 
\end{eqnarray}

For each mode $a$ we have simply obtained the total weight of 
trajectories contributing to transmission. In the classical limit
the sum over $a$ is converted into an integral over the initial angle
$\theta$ and we write

\be
\langle T \rangle = N \tau
\ee

\nin with

\be
\label{tracin}
\tau= \int_{0}^{\pi/2} d\theta \cos\theta \ \sum_{n} \frac{\delta_n}{W} \ .
\ee

Therefore, the total transmission coefficient is proportional to the 
number of modes, and the constant $\tau$ is a purely geometric factor.
Breaking the contribution of families into that of individual trajectories 
we are left with the usual classical transmission probability \cite{Chaos}

\begin{eqnarray}
     \tau = \frac{1}{2} \int_{-\pi/2}^{\pi/2} d\theta \cos\theta 
\int_0^W \frac{d y}{W} \ f(y,\theta) \ ,
 \label{trac}
\end{eqnarray}

\nin where $f(y,\theta)=1$ if the trajectory with initial
conditions $(y,\theta)$ is transmitted and $f(y,\theta)=0$ otherwise.

The mean slope in the numerical results of Fig.~\ref{fig:II} (with $W=0.2$) is $0.67$, 
in excellent agreement (better than $0.5\%$) with the classical transmission probability 
$\tau$. Also, the local mean $\langle T(k) \rangle$ of Eq.~(\ref{eq:simpspl2}) presents
the same slope $\tau$ and the mode quantization gives rise to fluctuations around
this mean (which remain much smaller than those of $T(k)$). The classical coefficient 
$\tau$ can be obtained, as in Eq.~(\ref{trac}), by sampling the space of classical 
trajectories with random choices of initial conditions $\theta$ and $y$ \cite{Chaos}, 
or more efficiently, by sampling the angles $\theta$ and incorporating the weights 
$\delta_n$ emerging from the intermediate fractions of $1/\tan\theta$, as 
suggested by Eq.~(\ref{tracin}). We then see 
that the continued fraction approach is not only useful for evaluating 
semiclassical effects, but also for classical properties like the 
transmission coefficient or the length distribution. Random sampling
of classical trajectories is an appropriate procedure for chaotic
structures, where the ergodicity of phase space results in an 
exponential distribution of lengths. On the other hand, integrable
cavities exhibit power-law distributions, which are more difficult
to obtain by trajectory sampling. In this case, the continued fraction
approach is very efficient since, for a given angle, only a finite number
of terms are relevant, and the contributing families are incorporated 
at once according to their weight.

The straight line that best approximates the transmission coefficient $T(k)$
is $\tau N + \kappa$. The
constant $\kappa$ is related with the elastic backscattering and also depends
on the underlying classical dynamics. In particular, the dependence of $\kappa$
on the magnetic field results in the weak localization effect \cite{bar,Chaos}.
However, we will not address the effect of a magnetic field in this work, nor 
the calculation of $\kappa$.


\section {Conductance Fluctuations}
\label{sec:coflu}

As visible from Fig.~\ref{fig:II}, the oscillations around the
mean transmission coefficient $\tau N + \kappa$ grow with larger $N$, consistently
with the numerical quantum mechanical calculations of Ref.~\cite{Chaos}. We will now
evaluate the local fluctuations, that is $\langle \delta T^2 \rangle =
\langle (T-(\tau N + \kappa))^2 \rangle$. We begin with the simplified
expression (\ref{eq:simp}) of $T$, and write
\begin{eqnarray}
  T^2 = \frac{1}{W^4} \sum_{a,b,a',b'=1}^{N} \frac{\cos \theta_b}{\cos
  \theta_a}\frac{\cos \theta_{b'}}{\cos \theta_{a'}} 
  \sum_{n,n^{\prime},n^{\prime\prime},n^{\prime\prime\prime}} 
\varepsilon_n \overline{\varepsilon}_{n^{\prime}} \varepsilon_{n^{\prime\prime}} 
\overline{\varepsilon}_{n^{\prime\prime\prime}} \nonumber \\
\times\varepsilon_n^{\prime} \overline{\varepsilon}_{n^{\prime}}^{\prime} 
\varepsilon_{n^{\prime\prime}}^{\prime} 
\overline{\varepsilon}_{n^{\prime\prime\prime}}^{\prime}
\delta_n \delta_{n^{\prime}} \delta_{n^{\prime\prime}} \delta_{n^{\prime\prime\prime}} 
\exp{\left[ ik (\tilde{L}_n\!-\!\tilde{L}_{n'}\!+\!\tilde{L}_{n^{\prime\prime}}\!-\!
\tilde{L}_{n^{\prime\prime\prime}})\right]}
\nonumber
\end{eqnarray} 
\vspace{-0.7cm}
\be
\label{eq:g2}
\times\Pi_n(a\!-\!b) \overline{\Pi}_{n^{\prime}}(a\!-\!b) \Pi_{n^{\prime\prime}}(a'\!-\!b')
  \overline{\Pi}_{n^{\prime\prime\prime}}(a'\!-\!b') \ .
\ee

If we take the average over several fluctuations, we only consider the 
terms having a null phase, i.e.
\begin{equation}
  \label{eq:ph0}
 \tilde{L}_n-\tilde{L}_{n'}+\tilde{L}_{n^{\prime\prime}}-
\tilde{L}_{n^{\prime\prime\prime}} = 0 \ .
\end{equation}

\nin This condition is satisfied with the pairing  $\tilde{L}_n=\tilde{L}_{n'}$ 
and $\tilde{L}_{n^{\prime\prime}}=\tilde{L}_{n^{\prime\prime\prime}}$, but the
resulting term cancels against the square of the average transmission coefficient. 
A non trivial pairing is obtained when $\tilde{L}_n=\tilde{L}_{n^{\prime\prime\prime}}$
and $\tilde{L}_{n'}=\tilde{L}_{n^{\prime\prime}}$ with $\tilde{L}_n \ne 
\tilde{L}_{n'}$, which implies $a\!=\!a'$. The contribution of this pairing to the
local fluctuations is 

\begin{eqnarray}
 \langle \delta T^2 \rangle_{I} = 
  \frac{1}{W^4}\sum_{a=1}^{N} \sum_{b,b'=1}^{N} 
  \frac{\cos \theta_b \cos \theta_{b'}}{\cos^2 \theta_a} \nonumber \\ 
  \times\sum_{n,n'} \Pi_n(a\!-\!b)\overline{\Pi}_{n'}(a\!-\!b)\Pi_{n'}(a\!-\!b')
  \overline{\Pi}_n(a\!-\!b') \delta_n^2 \delta_{n'}^2 = \nonumber \\
  \frac{1}{W^4}\sum_{a=1}^{N} \sum_{n,n'} \left(\min{\left\{\frac{W}{\delta_n},
\frac{W}{\delta_{n'}}\right\}}
\right)^2 \delta_n^2 \delta_{n'}^2 \ .
\label{eq:fluflu}
\end{eqnarray}

In the semiclassical limit the sum over $a$ can be converted into an integral
dictating a linear behavior of $\langle \delta T^2 \rangle_{I}$ with respect
to $k$

\be 
\label{eq:flufluf}
\langle \delta T^2 \rangle_{I} = \frac{N}{2} \
\int_{-\pi/2}^{\pi/2} d\theta \cos\theta \sum_{n,n'} 
\left(\frac{\min{\{\delta_n,\delta_n'\}}}{W} \right)^2 \ .
\ee
 
\nin The proportionality coefficient is only related to the geometry of the cavity, 
and can be computed using continued fractions.

The two pairings above described are those usually considered in dealing with chaotic
cavities, except that in such cases we take individual trajectories instead
families. In the integrable system we are studying there is another non trivial
way of satisfying Eq.~(\ref{eq:ph0}), that is, $\tilde{L}_n\!-\!\tilde{L}_{n'}=
\tilde{L}_{n^{\prime\prime}}\!-\!\tilde{L}_{n^{\prime\prime\prime}}$, with
$\tilde{L}_n \ne \tilde{L}_{n'}$ and $\tilde{L}_n \ne \tilde{L}_{n^{\prime\prime}}$.
This typically happens when $n,n',n^{\prime\prime}$, and $n^{\prime\prime\prime}$
belong to the same ($m$-th) Farey sequence and they are respectively associated
with the intermediate functions $(p_m^{k+j},q_m^{k+j})$, $(p_m^{k},q_m^{k})$,
$(p_m^{k'},q_m^{k'})$, and $(p_m^{k'+j},q_m^{k'+j})$, with $\!k\ne\!k'$, $j\!\ne\!0$
and $k,k',k\!+\!j,k'\!+\!j \in (0,a_{m})$. Also, since we are dealing with 
transmission coefficients, we need $p_m^{k}$, $p_m^{k'}$, $p_m^{k+j}$ and $p_m^{k'+j}$
to be odd, which implies that if $j$ is odd, $P_{m-1}$ must be even. Under such conditions, 
according to Eqs.~(\ref{eq:redlength}) and (\ref{eq:deffar}) we have

\be
\tilde{L}_n-\tilde{L}_{n'} = j \left(P_{m-1} \cos \theta_a + Q_{m-1} \sin \theta_a
\right) \ .
\ee

\nin If neither $n$ nor $n'$ correspond to the first family of the sequence
($k,k' \ne 0$), Eqs.~(\ref{exlc}), (\ref{exuc}) and (\ref{eq:deffar}) dictate
that the four exiting intervals have the same length

\be
\delta_m= \left|P_{m-1}\tan \theta - Q_{m-1}\right| \ .
\ee

This last contribution to the local fluctuations can be expressed as a sum
over the convergents

\be
  \langle \delta T^2 \rangle_{II} = \sum_{a=1}^{N} \sum_{m}
  \sum_{k,k'=0}^{a_m} \sum_{j} \frac{\delta_m^2}{W^2} \ ,
\ee

\nin with the above specified restrictions for $k,k'$ and $j$. As in the 
previous case, converting the sum over $a$ into an integral yields a
contribution $\langle \delta T^2 \rangle_{II}$ to the local fluctuations
that is linear in $k$, with a purely geometrical coefficient given by the
continued fraction representation of $1/\tan \theta$.

The linearity of $\langle \delta T^2 \rangle=\langle \delta T^2 \rangle_{I}+
\langle \delta T^2 \rangle_{II}$ with $k$ that we have demonstrated is in
good agreement with our numerical results of Fig.~\ref{fig:II}. In the inset
we show the numerically obtained $\langle \delta T^2 \rangle$ and the
straight line representing $\langle \delta T^2 \rangle_{I}$ from 
Eq.~(\ref {eq:flufluf}), with a proportionality coefficient of $0.13$ calculated
from the continued fractions. The agreement is very good and shows that 
$\langle \delta T^2 \rangle_{II}$ is negligible in comparison with 
$\langle \delta T^2 \rangle_{I}$. Such behavior is understandable since, for 
a given incoming mode $a$, the latter is given by a double sum over families
of trajectories ($n$ and $n'$), while the former is given as a [single]
sum over the convergents ($m$) and sums within the associated Farey
sequence, provided it verifies the above established conditions. The denominators
$Q_m$ of the convergents increase exponentially with $m$, $Q_m \ge 2^{(m-1)/2}$ \cite{khinchin},
or more precisely as Greenman has recently proved \cite{greenman} 

\be
\lim_{m\rightarrow \infty} Q_m^{1/m} = \exp{\left[\frac{\pi^2}{12 \log 2}\right]} \ ,
\ee

\nin bar a set of measure zero. Therefore, the $a_m$ of Eqs.~(\ref{eq:khin1}) and 
(\ref{eq:khin2}), giving the number of 
intermediate fractions of the $m$-th sequence, remain of order $1$ even in the
large $m$-limit. Consequently, there are much more terms contributing to 
$\langle \delta T^2 \rangle_{I}$ than to $\langle \delta T^2 \rangle_{II}$.

In chaotic cavities the diagonal approximation is not enough to describe the
magnitude of the conductance fluctuations, due to the exponential proliferation
of trajectories that results in almost degenerate actions. This is clearly not 
the case in rectangular cavities, where we have shown the good agreement between
$\langle \delta T^2 \rangle_{I}$ and the numerical results.

\begin{figure}[tbh]
\setlength{\unitlength}{1mm}
\begin{picture}(10,53)(85,10)
\put(0,0){\epsfxsize=7cm\epsfbox{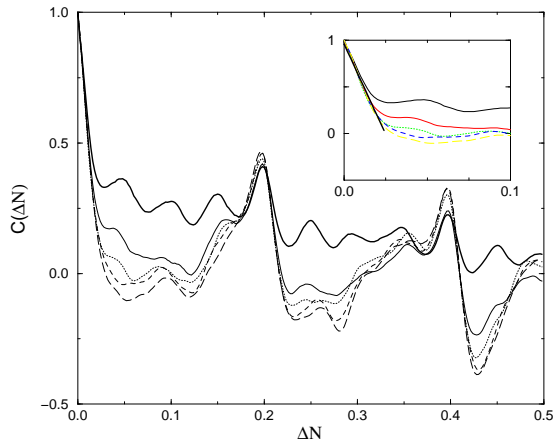}}
\end{picture}
\vspace{4mm}
\caption{Correlation function locally normalized according to 
Eq.~(\ref{eq:corrfun}) for five $N$ intervals: 0-20 (thick solid),
20-40 (thin solid), 40-60 (dotted), 60-80 (dashed) and 80-100 (long-dashed).
Inset: blow up of the small $\Delta N$ region showing a triangular 
behavior at the origin for all the intervals considered. }
\label{fig:III}
\end{figure}

Conductance fluctuations are usually characterized by their amplitude
$\langle \delta T^2 \rangle$ and their energy-correlation length. In chaotic
cavities, random matrix theory and numerical simulations \cite{meba,ropibe2}
yield an universal $\langle \delta T^2 \rangle$, independent of energy. The
correlation length can be obtained from a semiclassical approach 
\cite{blusm,jal} and is proportional to the escape rate (the inverse characteristic
time that scattering trajectories spend in the cavity). In our integrable cavity
we have seen that $\langle \delta T^2 \rangle$ is not universal, but energy
dependent. Also, there does not exist a characteristic time for exiting the
cavity. Therefore it is not obvious that a correlation function depending 
only on the energy increment can be defined. That is why we consider the
correlation function 

\begin{equation}
\label{eq:corrfun}
  C(\Delta k) = \frac{\langle \delta T(k\!+\!\Delta k)\delta
  T(k)\rangle}{\langle \delta  T^2(k) \rangle} \ ,
\end{equation}

\nin where $k$ varies on an interval much larger than $\Delta k$, but
small enough to neglect secular variations. 

From Eqs.~(\ref{eq:corrfun})
and (\ref{eq:tab}) we obtain numerically the correlation functions
shown in Fig.~\ref{fig:III} for various $N$ intervals from 0-20 (thick solid)
to 80-100 (long-dashed).
Except for the first interval, where the semiclassical approximation is most
questionable, the form of $C(\Delta k)$ seems to be almost independent of
the region of evaluation. Moreover, a singularity for small $\Delta k$
appears in all the $N$-intervals in the form of a cusp around the origin (inset).
The linear behavior of $C(\Delta k)$ is to be contrasted to the lorentzian
correlation function expected for a chaotic cavity. This difference between
integrable and chaotic cavities is reminiscent of what happens with the weak
localization effect: a semiclassical approach predicts a linear magneto resistance
for integrable cavities and a lorentzian lineshape for chaotic ones \cite{bar}.
The linear behavior of the correlation function is consistent with the quantum
calculations of Ref.~\cite{Chaos} that yielded a decay of the power spectrum
(Fourier transform of $C(\Delta k)$) as $x^{-2}$ at large $x$.

The oscillatory structure of $C(\Delta N)$ on the scale of $\Delta N \simeq 0.2$
corresponds to a typical length of twice the size of the square. We can
understand this behavior by noticing that, for a given angle $\theta$, 
the smallest possible difference between the length of two trajectories is 2.
The analytical calculation of $C(\Delta k)$ is considerably more involved than that of
$\langle \delta T^2 \rangle$; we will have a sum as in Eq.~(\ref{eq:fluflu}),
where each term should be multiplied by $\exp{[i\Delta k(\tilde{L}_n-\tilde{L}_{n'})]}$.
Therefore it looks rather difficult to extract analytically the triangular behavior around 
the origin.


\section {Conclusions}
\label{sec:conclu}

In this work we have developed a semiclassical method to calculate the
transmission through a rectangular cavity, based on continued fractions.
We have proved that the expansions, over families of trajectories, are
finite and reproduce the basic features of quantum mechanical 
calculations. The conductance fluctuations, within our semiclassical 
approximation, are shown to be qualitatively
different from the chaotic case: they increase linearly with the incoming
Fermi wave-vector (with a proportionality coefficient that can be easily
obtained with our scheme) and the correlation function presents a 
cusp-like singularity at the origin.
The continued fraction approach is also useful to address classical 
properties like the transmission coefficient, the length distribution,
or the area distribution (responsible for the the shape and magnetic
field scale of the weak-localization effect). 

We have seen that open systems exhibit larger fluctuations
when the classical dynamics is integrable. The situation is analogous to the
density of states of closed systems, which is characterized by stronger
fluctuations in integrable than in chaotic geometries. The augmented 
fluctuations in {\em integrable closed and open} geometries can be
traced to the same origin: {\em the bunching of trajectories into
families} in the semiclassical expansions, the Berry-Tabor formula and 
Eq.~(\ref{eq:tab}) respectively.

In the example discussed in this work the classical dynamics in the leads
has the same conserved quantities than in the cavity, leading to bunching
of trajectories in families of degenerate action. This is not the case in
circular cavities, where a semiclassical expansion based on individual
trajectories has been developed \cite{Lin}. Therefore, the geometry of the
leads plays a very important role \cite{bird} and renders the quantum 
signatures of integrability in open systems quite involved.

The existing experimental results \cite{marcu,lee97} show the difference in
the conductance fluctuations of integrable and chaotic cavities, but more work
should be undertaken to test our results. The variation of Fermi energy
(or number of modes $N$) has been achieved in chaotic and integrable ballistic
microstructures \cite{keller,zozou97}. However, the intervals $\Delta N$ of
variation remain much smaller than those allowing to detect a linear 
increase of the variance. Therefore, microwave cavities \cite{stoec} appear
as more promising for the observation of such an effect.

\acknowledgements
We are indebted to H.~Baranger, K.~Richter and D.~Ullmo for important remarks on the
manuscript. We thank J.~Bellisard and J.~Keating for helpful hints on continued fractions.

\end{multicols}


\begin{references}
\bibitem{RevBoh} {\it Chaos and Quantum Physics}, edited by M.-J.~Giannoni, 
A.~Voros, and J.~Zinn-Justin (North-Holland, Amsterdam, 1991).
\bibitem{BGS}O.~Bohigas, M.J.~Giannoni, and C.~Schmit,
Phys. Rev. Lett. {\bf 52}, 1 (1984); O.~Bohigas in Ref.~\cite{RevBoh}, p~87.
\bibitem{gutz90} M.C.~Gutzwiller, {\it Chaos in Classical and Quantum
Mechanics} (Springer, Berlin, 1990); and in Ref.~\cite{RevBoh}, p~201.
\bibitem{ber77} M.V.~Berry and M.~Tabor, J.~Phys.~A {\bf 10} 371 (1977).
\bibitem{Miller} W.H.~Miller, Adv. Chem. Phys. {\bf 25} 69 (1974).
\bibitem{jal} R.A.~Jalabert, H.U.~Baranger, and A.D.~Stone,
Phys.\ Rev.\ Lett.\ {\bf 65}, 2442 (1990). 
\bibitem{Chaos} H.U.~Baranger, R.A.~Jalabert, and A.D.~Stone, Chaos, {\bf 3},
665 (1993).
\bibitem{ishio95} H.\ Ishio and J.\ Burgd\"orfer, Phys.\ Rev.\ B {\bf 51},
2013 (1995).
\bibitem{Lin} W.A.~Lin and R.~Jensen, Phys.\ Rev.\ B {\bf 53}, 3638 (1996);
 W.A.~Lin, Chaos, Solitons\&Fractals {\bf 8}, 995 (1997).
\bibitem{schwi96} C.\ D.\ Schwieters, J.\ A.\ Alford, and J.\ B.\ Delos,
Phys.\ Rev.\ B {\bf 54}, 10652 (1996).
\bibitem{Wirtz} L.~Wirtz, J.-Z.~Tang, and J.~Burgd\"orfer, Phys.\ Rev.\ B {\bf 56},
7589 (1997).
\bibitem{marcu} C.M.~Marcus {\em et al.} Phys.\ Rev.\ Lett.\ {\bf 69}, 506 (1992);
and Chaos {\bf 3}, 643 (1993).
\bibitem{timp} {\it Nanotechnology}, edited by B.~Timp (AIP Press, 1995).
\bibitem{csf} {\it Chaos and Quantum Transport in Mesoscopic Cosmos}, special issue,
Chaos, Solitons\&Fractals {\bf 8}, 971-1412 (1997).
\bibitem{stoec} H.-J.\ St\"ockmann and J.\ Stein, Phys.\ Rev.\ Lett.\ {\bf 64},
2215 (1990); H.-D.\ Gr\"af {\em et al.} Phys.\ Rev.\ Lett.\ {\bf 69}, 1296 (1992);
M.\ Kollmann {\em et al.} Phys.\ Rev.\ E {\bf 49}, R1 (1994).
\bibitem{keller} M.W.~Keller {\em et al.} Surf.\ Sci.\ {\bf 305}, 501 (1994); and
Phys.\ Rev.\ B {\bf 53}, R1693 (1996).
\bibitem{lee97} Y.\ Lee, G.\ Faini, and D.\ Mailly, Phys.\ Rev.\ B, {\bf 56}, 9805 
(1997); and in Ref.~\cite{csf}, p.~1325.
\bibitem{blusm} R.~Bl\"umel and U.~Smilansky, Phys. Rev. Lett.
{\bf 64}, 241 (1990).
\bibitem{meba} H.U.~Baranger and P.A.~Mello, Phys. Rev. Lett.
{\bf 73}, 142 (1994); H.U.~Baranger in Ref.~\cite{timp}.
\bibitem{ropibe2} R.A.~Jalabert, J.-L.~Pichard, and C.W.J.~Beenakker,
Europhys. Lett. {\bf 27}, 255 (1994); C.W.J.~Beenakker, Rev. of Mod. Phys. {\bf 69},
731 (1997).
\bibitem{huibers} A.G.~Huibers {\em et al.} Phys. Rev. Lett. {\bf 81}, 1917 (1998); 
C.M.~Marcus {\em et al.}, in Ref.~\cite{csf}, p.~1261.
\bibitem{bar} H.U.~Baranger, R.A.~Jalabert, and A.D.~Stone,
Phys.\ Rev.\ Lett.\ {\bf 70}, 3876 (1993).
\bibitem{chang94} A.M.~Chang, H.U.~Baranger, L.N.~Pfeiffer, and
K.W.~West, Phys. Rev. Lett. {\bf 73}, 2111 (1994); A.M.~Chang,
in Ref.~\cite{csf}, p.~1281.
\bibitem{berry94} M.J.~Berry, J.H.~Baskey, R.M.~Westervelt, and
A.C.~Gossard, Phys.\ Rev.\ B {\bf 50}, 8857 (1994).
\bibitem{lutj} G.~L\"utjering {\em et al.} Surf. Sci. {\bf 361/362}, 709 (1996).
\bibitem{bird} J.P.~Bird {\em et al.},  Phys.\ Rev.\ B {\bf 52}, R14336 (1995);
and in Ref.~\cite{csf}, p.~1299.
\bibitem{zwanzig} R.~Zwanzig, J.~Stat.~Phys. {\bf 30}, 255 (1983).
\bibitem{fish81} D.S.~Fisher and P.A.~Lee, Phys.\ Rev.\ B {\bf 23},
6851 (1981).
\bibitem{ingold} M.~Schreier, K.~Richter, G.-L.~Ingold, and R.A.~Jalabert,
Eur. Phys. J. B {\bf 3}, 387 (1998).
\bibitem{khinchin} A.Ya.~Khinchin, {\em Continued Fractions} (University of
Chicago Press, Chicago, 1964).
\bibitem{greenman} C.~Greenman, J.~Phys. A {\bf 30}, 915 (1997).
\bibitem{zozou97} I.\ V.\ Zozoulenko, R.\ Schuster, K.-F.\ Berggren, and K.\
Ensslin, Phys.\ Rev.\ B {\bf 55}, R10209 (1997).

\end{references}
\end{document}